\documentstyle[elsevier,12pt,epsfig]{article}

\begin{document}

\newlength{\htwid}
\setlength{\htwid}{0.48\textwidth}
\newlength{\fvskip}
\setlength{\fvskip}{-9pt}
\newlength{\cvskip}
\setlength{\cvskip}{-15pt}

\newcommand{\be}{\begin{equation}}
\newcommand{\ee}{\end{equation}}
\newcommand{\qq}{$\langle \bar q q \rangle $}
\newcommand{\rmp}[1]{{ Rev.\ Mod.\ Phys.\ }{\bf A{#1}}}
\newcommand{\npa}[1]{{ Nucl.\ Phys.\ }{\bf A{#1}}}
\newcommand{\npb}[1]{{ Nucl.\ Phys.\ }{\bf B {#1}}}
\newcommand{\prc}[1]{{ Phys.\ Rev.\ }{\bf C {#1}}}
\newcommand{\prd}[1]{{ Phys.\ Rev.\ }{\bf D{#1}}}
\newcommand{\plb}[1]{{ Phys.\ Lett.\ }{\bf{#1}B}}
\newcommand{\prl}[1]{{ Phys.\ Rev.\ Lett.\ }{\bf {#1}}}
\newcommand{\aop}[1]{{ Ann.\ Phys.\ }{\bf {#1}}}
\newcommand{\zpa}[1]{{ Z.\ Phys.\ }{\bf A {#1}}}
\newcommand{\zpc}[1]{{ Z.\ Phys.\ }{\bf C {#1}}}
\newcommand{\pr}[1]{{ Phys.\ Rep.\ }{\bf {#1}}}
\newcommand{\phr}[1]{{ Phys.\ Rep.\ }{\bf {#1}}}
\newcommand{\varro}{\rho}

\vspace*{2cm}
\title{Signatures of dense hadronic matter %and dilepton production
in 
ultrarelativistic heavy ion reactions\footnote
{Supported by BMBF, DFG and GSI, ${}^b$E-mail:
ernst@th.physik.uni-frankfurt.de, 
${}^c$Invited speaker
}}
\author{L.A.\ Winckelmann${}^{d}$, 
C.\ Ernst${}^{b,d}$, L.\ Gerland${}^d$, J.\ Konopka${}^d$, S.\ Soff${}^e$, 
S.A.\ Bass${}^{d}$, M.\ Bleicher${}^d$, M.\ Brandstetter${}^d$, A.\ Dumitru${}^d$, 
C.\ Spieles${}^d$, H.\ Weber${}^d$,
C.\ Hartnack${}^f$, J.\ Aichelin${}^f$, N.\ Amelin${}^g$,
H.\ St\"ocker${}^{cd}$ and W.\ Greiner${}^d$}
\address{${}^d$Institut f\"ur Theoretische Physik,
         Johann Wolfgang Goethe-Universit\"at, Frankfurt, Germany 
\\
${}^e$Gesellschaft f\"ur Schwerionenforschung, 
Darmstadt, Germany \\
${}^f$SUBATECH, Ecole des Mines,
Nantes,  France\\
${}^g$Joint Institute for Nuclear Research (JINR),
Dubna, Russia}

\maketitle\abstracts{
\noindent
{\bf Abstract:} The behavior of hadronic matter at high baryon densities is studied
within  Ultrarelativistic Quantum Molecular Dynamics (URQMD). 
Baryonic stopping is observed for Au+Au collisions from SIS up to
SPS energies. 
The 
excitation function of % hadronic abundances, stopping and 
flow shows strong sensitivities to the underlying equation of state (EOS), 
allowing for  systematic studies of the EOS. 
Dilepton spectra are calculated with and without shifting the $\rho$ pole.
Except for S+Au collisions our calculations reproduce the CERES data.
%Effects of a  density  
%dependent pole of the $\rho$-meson propagator on
%dilepton spectra are studied for different systems and centralities at 
%CERN energies.
}

\section{Introduction}

The only possibility to probe excited nuclear matter in the
laboratory are nucleus--nucleus reactions \cite{Sto86}.
In particular when two heavy ions
like Au or Pb collide most centrally, the combined system forms a zone of
high (energy) density and high excitation of the involved constituents.
The transient pressure at high density has specific
dynamic implications, such as collective sideward flow.
Hence, fundamental properties like the repulsion of the
nuclear  equation of state (EOS)
are studied
via event shape analysis of nucleons and clusters
\cite{Dos86,matt95,Part95}.
The EOS at fixed temperature %can be interpreted as
%interpreted microscopically 
yields
a density dependent potential %modifiying the nucleon mass.
and a modified nucleon mass.
At low densities these effects %is similaras has been 
are proposed by  % various calculations on 
chiral %limits of the Skyrme 
lagrangians\cite{GEB91}.
%the constituent quark mass\cite{KKu92} and
%the chiral condensate \qq %, i.e.\ the order parameter  of chiral smmetry
\cite{THa92a}.
Since the chiral condensate \qq{} relates  closely to hadron masses,
the decay of short lived vector mesons,
observed through the dilepton channel, 
is suggested as a promising
experimental signal
to investigate the gradual restoration of chiral symmetry.

\section{Ultrarelativistic Quantum Molecular Dynamics}

\begin{figure}[t]
\newcommand{\mysm}{\xpt}
\newcommand{\myno}{\xipt}
\begin{minipage}[c]{1.16\htwid}
\setlength{\unitlength}{3pt}
{\mysm
\begin{picture}(88,68)
\put(0,30.3){{\mysm
\begin{tabular}{cccccc}
 \hline  \hline  %\\[-9pt]
{\myno N}&{\myno  $\Delta$}&{\myno $\Lambda$}&{\myno $\Sigma$}
&{\myno $\Xi$}&{\myno  $\Omega$} \mysm \\  \hline
 ${938}$&$ {1232}$&  ${1116}$& ${1192}$& ${1317}$& ${1672}$\\
 ${1440}$& ${1600}$& ${1405}$& ${1385}$& ${1530}$&\\
 ${1520}$& ${1620}$& ${1520}$& ${1660}$& ${1690}$&\\
 ${1535}$& ${1700}$& ${1600}$& ${1670}$& ${1820}$&\\
 ${1650}$& ${1900}$& ${1670}$& ${1790}$& ${1950}$&\\
 ${1675}$& ${1905}$& ${1690}$& ${1775}$&$$&\\
 ${1680}$& ${1910}$& ${1800}$& ${1915}$&$$&\\
 ${1700}$& ${1920}$& ${1810}$& ${1940}$&$$&\\
 ${1710}$& ${1930}$& ${1820}$& ${2030}$&&\\
 ${1720}$& ${1950}$& ${1830}$&$$&&\\
 ${1990}$&         & ${2100}$&$$&&\\
&& ${2110}$&$$&&\\
\hline\hline
\end{tabular}}}
\put(42,11){{\mysm
\begin{tabular}{ccccc}
\hline \hline %\\[-9pt]
$0^-$ & $1^-$ &$ 0^+$ &$ 1^+$ &$ 2^+$ \\ \hline
 $\pi$ & $ \rho$ & $ a_0$ & $ a_1$ & $ a_2$ \\
 $K  $ &$   K^*$ & $ K_0^*$ & $ K_1^*$ & $ K_2^*$\\
 $\eta$&$  \omega$& $ f_0 $&  $f_1$ & $ f_2 $\\
 $\eta'$&  $\phi $&  $\sigma$ & $ f_1'$& $ f_2'$\\
\hline\hline
\end{tabular}}}
\put(55,37){{\mysm
\begin{tabular}{c}
\hline \hline %\\[-9pt]
$ 1^-$ \\ \hline
$ \rho(1450)$ \\
$ \rho(1700)$\\
$ \omega(1420)$\\
$ \omega(1600)$\\
\end{tabular}}}
\end{picture}
}\\[9pt]
%}{tabelle}
%\end{center}
%\caption{\label{tabelle}
Table 1:
List of implemented baryons, mesons and their
resonances.
In addition {\em all} charge conjugate and iso-spin projected
states (and photons) are taken in and
treated on the same footing. 
\end{minipage}
\hfill
\begin{minipage}[c]{0.845\htwid}
\centerline{\psfig{figure=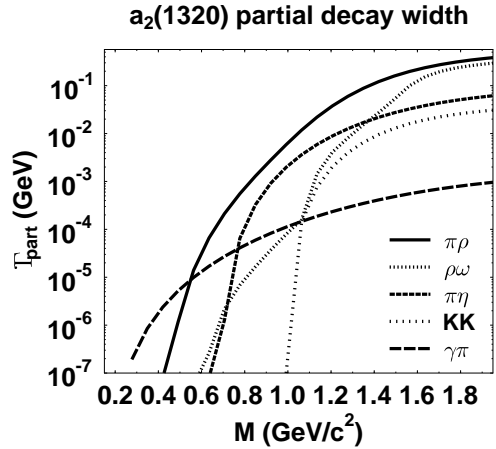,width=0.845\htwid}}
\caption{\label{fig:pwid}
$a_2$ partial decay rates into  specific channels.
The  average lifetime is given by the inverse of the sum.
Hence, in URQMD particles below resonance mass 
live longer,  due to shrinking phase space.}
\end{minipage}
\vspace{\fvskip}\end{figure}

Since many important aspects of nuclear matter
are not observable, numerical transport models are  suited
to test which assumptions are compatible to nature.
The present model (URQMD) \cite{uqmd,law96}
includes explicitely 50 different baryon species
(nucleon, delta, hyperon and their resonances up to  masses of 2.11~GeV)
and 25 different meson species (including strange meson resonances), which
are supplemented by % their corresponding antiparticles and
all isospin-projected states (see Table 1).
Symmetries regarding
time inversion,  iso-spin,
charge conjugation, etc.\
are implemented in a general manner,
e.g.\  all corresponding antiparticles are included and treated
on the very same (charge-conjugate) footing.
For excitations of higher masses a newly   developed 
string model is invoked.
It consistently allows for the population
of {\em all} included hadrons from a decaying string.
At low energies the dominant part of MM and MB
interactions is modeled via
$s$-channel reactions (formation and decays of resonances),
whereas  BB interactions are designed as exchange
of charge, strangeness and four momentum in the
$t$-channel.
For all resonances we use mass-dependent decay widths as illustrated in
Fig.\ref{fig:pwid} for the $a_2$ meson. The lifetime of resonances is
calculated as their inverse width. There exist, however, recent theoretical
ansatzes which yield a different mass dependence for the life-times of
resonances \cite{pawel}.
The real part of the baryon optical potential is modeled
according to the simple Skyrme ansatz, including Yukawa
and Coulomb forces.

\section{Creation of dense nuclear matter: stopping}

\begin{figure}[htb]
\begin{minipage}[t]{\htwid}
\centerline{\epsfig{figure=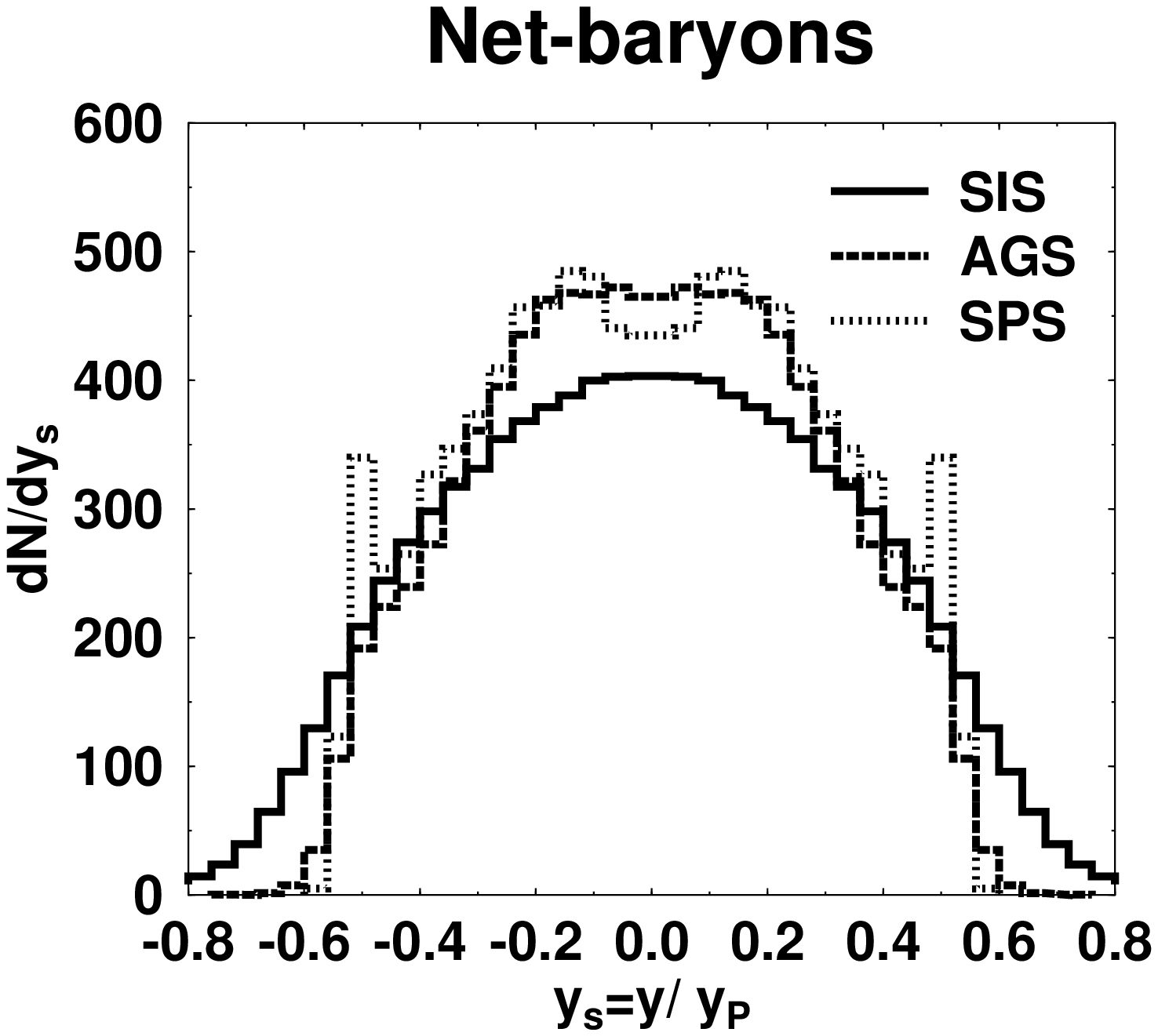,width=\htwid}} % stopp_bw.eps
\vspace{\cvskip}\caption{\label{fig:stopp}Rapidity distributions for Au+Au
collisions at SIS (1~$A$GeV),
AGS (10.6~$A$GeV) and Pb+Pb at CERN/SPS energies (160~$A$GeV).
%All distributions have been scaled to the projectile rapidity in the
%center of mass frame. 
}
\end{minipage}
\hfill
\begin{minipage}[t]{\htwid}
\centerline{\epsfig{figure=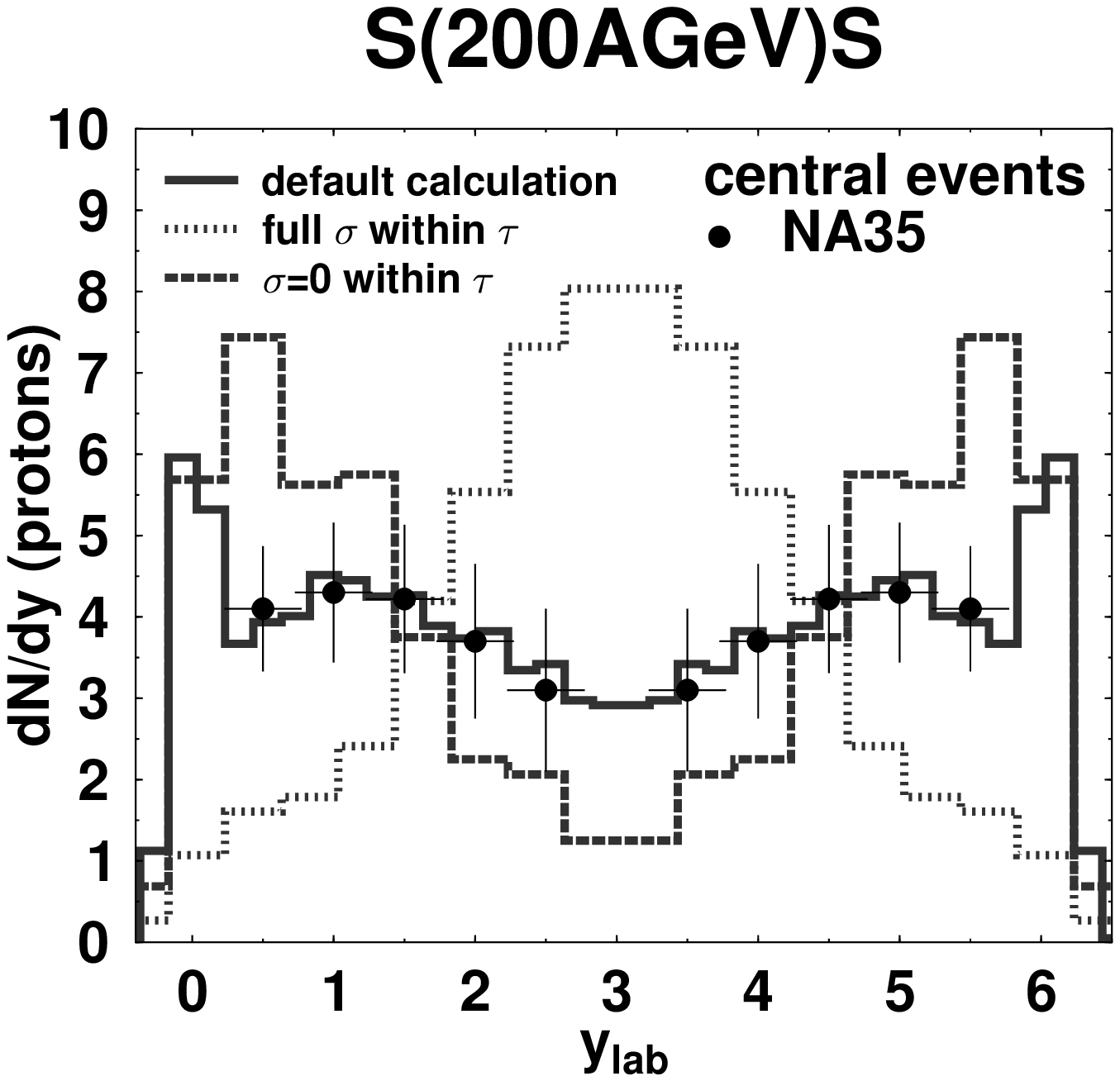,width=\htwid}}
\vspace{\cvskip}\caption{\label{fig:dndyform}
Rapidity distribution 
for S(200~$A$GeV)S for various treatments of the 
constituent (di-)quark cross section (see text). }
\end{minipage}
\vspace{\fvskip}
\end{figure}

Baryonic stopping is a necessary condition for the creation of hot and dense
nuclear matter. The key observable is the rapidity distribution of baryons.
It is displayed in  Fig.\ref{fig:stopp} and \ref{fig:pbpb} 
for heavy systems
such as Au+Au and Pb+Pb at energies referring to
three presently used heavy ion accelerators.
 %the system as heavy as Au+Au or Pb+Pb
In all cases gaussian rapidity distributions with peak around  
midrapidity %$y_s \sim \pm 0.2$
are found.
However, the physical processes associated
%(although induced by multiple collisions of the initial constituents)
show characteristic differences: The average longitudinal
momentum loss in the SIS energy regime is mainly due to the creation of
transverse momentum, whereas at  AGS/SPS energies abundant
particle production
%eats up
consumes a considerable amount of the incident beam energy.

At CERN/SPS energies baryon stopping is influenced
%determined not only by rescattering but
also by the formation time of strings which are
excited in  hard collisions.
In URQMD  baryons originating from a
leading constituent (di-)quark at the string edges
interact with (2/3)1/3
and mesons with 1/2 of their full cross sections during 
their formation time $\tau$.
The sensitivity on this reduction
is shown in Fig.\ref{fig:dndyform}
for the system S+S at 200~$A$GeV. The default calculation (including
formation time) reproduces the data  \cite{na35} 
fairly well whereas the calculation
with zero formation time (dotted line) exhibits strongest stopping.
A calculation with zero cross section within the formation time
gives transparency.

In order to study the influence of this this effect more
closely the $\sqrt{s}$ distributions
for Au+Au collisions at AGS and S+S collisions at SPS energies are
analyzed. Fig.\ref{fig:srtboth} (right) shows the respective distribution for
Au+Au. The collision spectrum is
dominated by BB collisons with full cross sections and exhibits a
maximum at low energies. Approximately 20\% of the collisions
involve a diquark, i.e. a baryon originating from a string decay whose
cross section is reduced to 2/3  of its full cross section.

%\begin{figure}[htb]
\begin{minipage}[t]{\htwid}
\centerline{\epsfig{figure=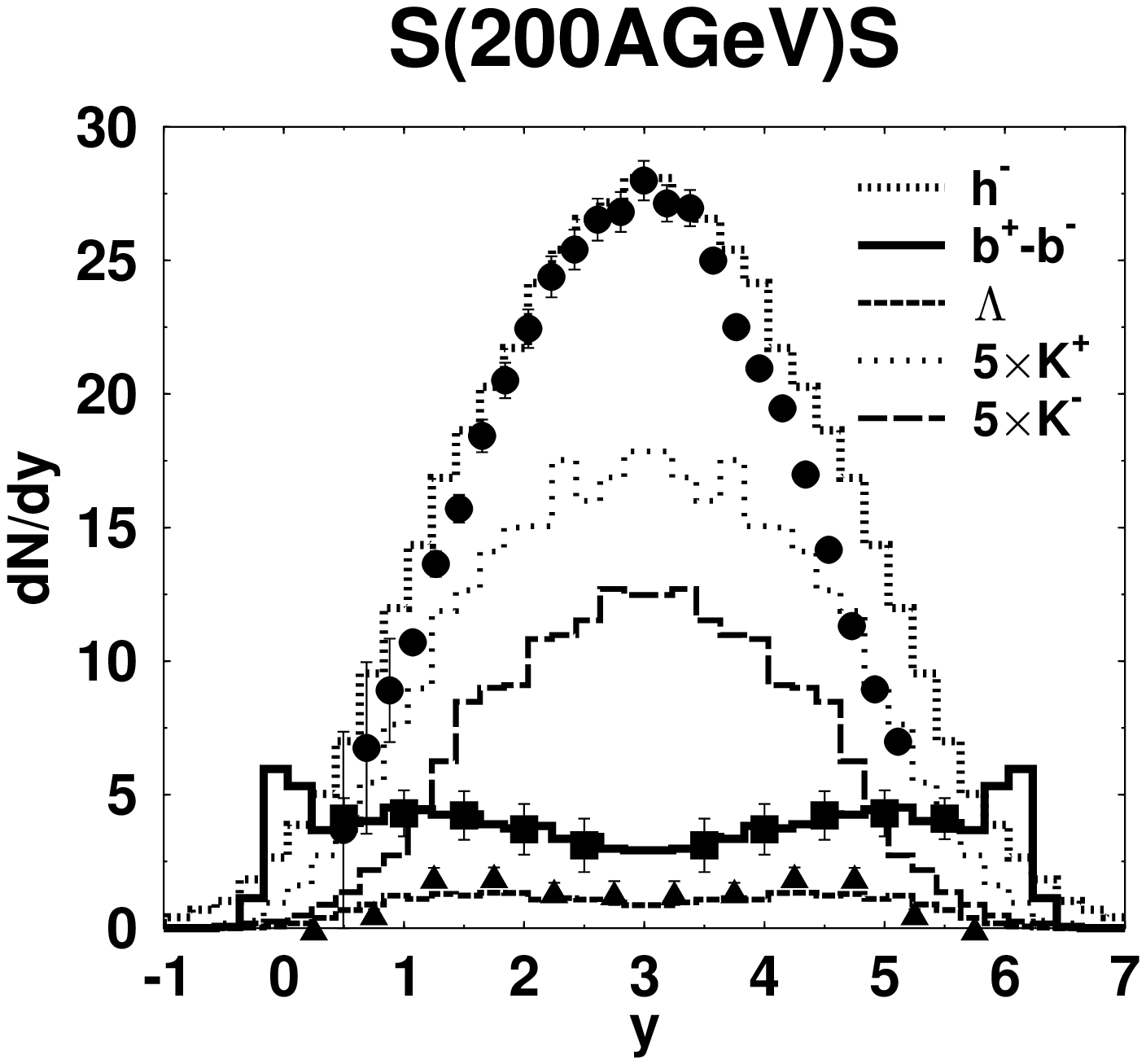,width=\htwid}} % stopp_bw.eps
%\vspace{\cvskip}
\end{minipage}
\hfill
\begin{minipage}[t]{\htwid}
\centerline{\epsfig{figure=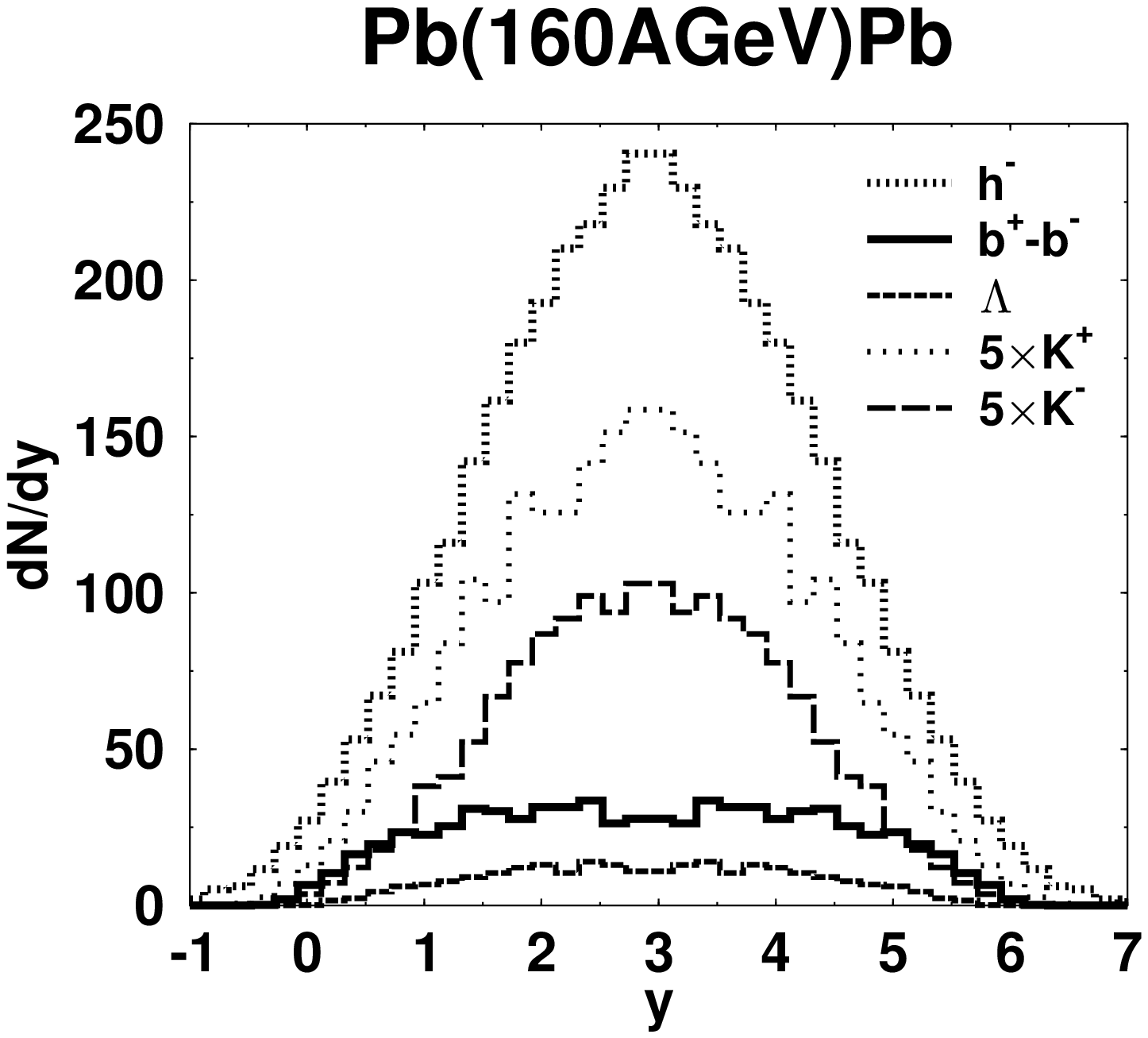,width=\htwid}}
%\vspace{\cvskip}
\end{minipage}
%\vspace{\fvskip}
%\caption{
\vspace{0.5\cvskip}

\figlabel{fig:pbpb} 
Rapidity distributions for S+S  at 200~$A$GeV (left) and 
Pb+Pb at 160~$A$GeV (right).
The  histograms label from top to bottom: negative
hadrons  ($h^-$),   kaons ($K^-$,  $K^+$),  protons ($b^+-b^-$) and lambdas
($ \Lambda $).  The kaons are multiplied by five. The symbols show 
data from  NA35 \cite{na35}.
%}
%\caption{
%\\ \figlabel{fig:pbpb} (right)
%{Rapidity distributions for Pb+Pb at 160~$A$GeV.
%The  histograms are labeled as in Fig.\ref{fig:s200s}. 
%}
%\end{figure}

In Fig.\ref{fig:srtboth} (left) the same analysis is performed for S+S at
200~$A$GeV. In contrast to the heavy system at AGS
the collision spectrum
exhibits two pronounced peaks dominated by full BB collisions,
one in the beam energy range and one in the low (thermal) energy  range.
Now approximately 50\% of the collisions, most of them at intermediate
$\sqrt{s}$ values, involve baryons stemming from string
excitations whose
cross sections are reduced by factors of 2/3  (referred to as
{\em diquarks}) or 1/3  (referred to as {\em quarks}).
The peak at high $\sqrt{s}$ values stems from the initial hard collisions
whereas the peak at low energies is related to the late, thermal stages
of the reaction.
%Since the reduction factor of 1/3 (2/3) does only take valence-quarks
%into account but neglects contributions of gluons and sea-quarks it is
%no undisputable quantity and contains a certain level of freedom -- the
%same holds true for the formation time itself which cannot be derived
%from first principles.
%%%%%%%%%%%%%%%%%%%%%%%%%%%%%%%%%%%%%%%%%%%
%In order to study this effect more closely the $\sqrt{s}$ distribution for 
%Au+Au collisions at AGS and 
%S+S reactions at SPS energies %are
%analyzed. Figure \ref{fig:srtboth} shows the respective distribution for
%Au+Au. The collision spectrum is
%dominated by BB collisons with full cross sections and exhibits a
%maximum at low energies. Approximately 20\% of the collisions
%involve di-quarks, i.e. leading baryons whose
%cross section are  reduced to 2/3  of their full cross sections
%during formation time.
%is shown 
%%In 
%in Fig.\ref{fig:srtboth}.  %the same analysis is performed 
%%for S+S at 200~$A$GeV. In contrast to the heavy system at AGS
%The collision spectrum
%exhibits two pronounced peaks dominated by  BB collisions,
%one in the beam energy range and one in the low (thermal) energy  range.
%Approximately 50\% of the collisions, most of them around 
%$ \sim 10 \pm 5~$GeV, involve
%%constituent quarks
%baryons
%during their formation time,
%%stemming from string excitations 
%whose cross sections are reduced by factors of 2/3
%(referred to as {\em di-quarks} )
%or 1/3  (referred to as {\em quarks}).
%This reduction leads to more transparency due to less collisions
%as compared to a calculation without reduction 
%(Fig.\ref{fig:dndyform}). 

\vspace{10pt}\noindent
%\begin{figure}[htb]
%\begin{minipage}[t]{\htwid}
\centerline{\epsfig{figure=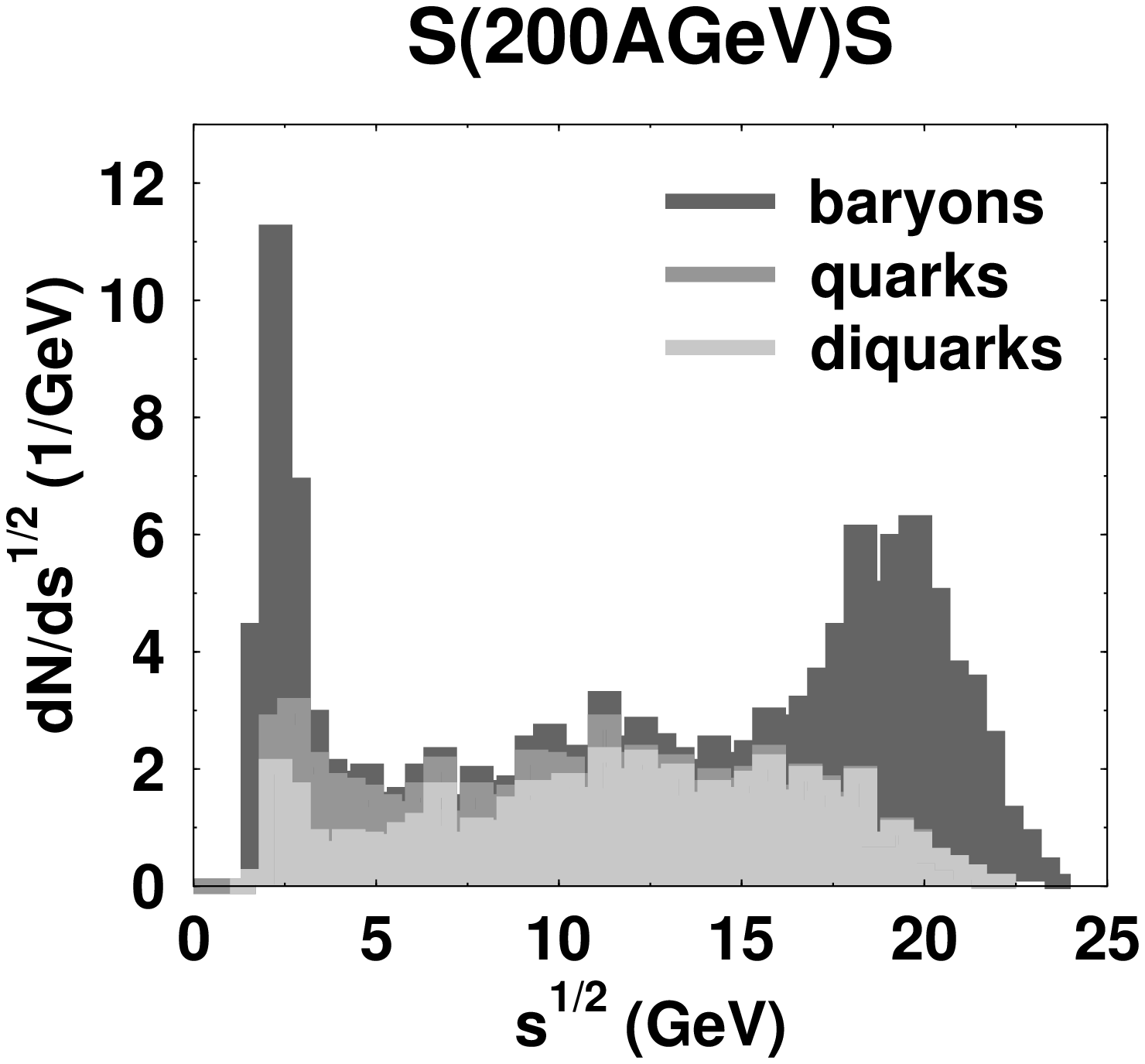,width=\htwid}\hfill
%\end{minipage}
 %\begin{minipage}[t]{\htwid}
%\centerline{
\epsfig{figure=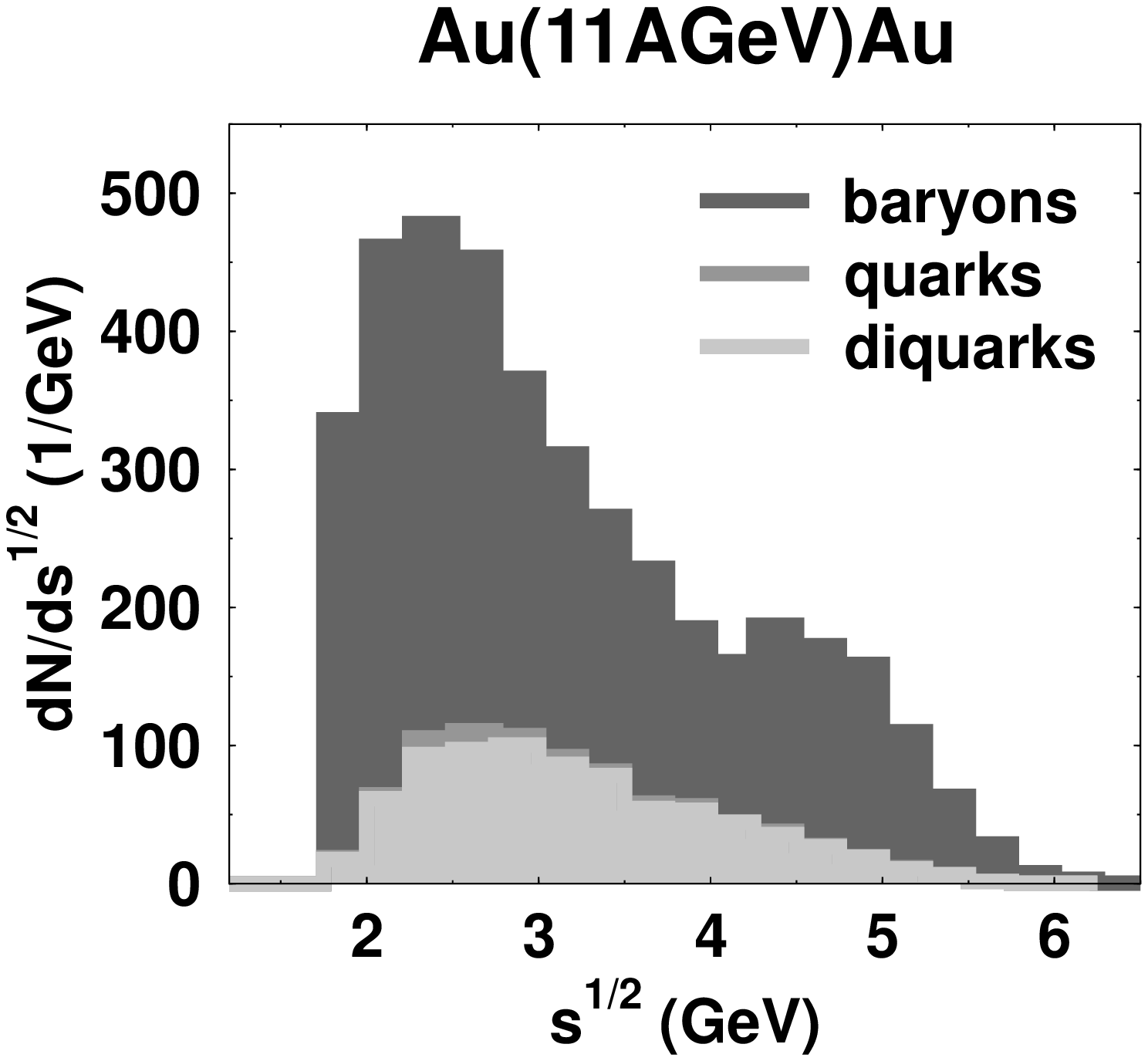,width=\htwid}}
%\vspace{0.5\cvskip}\caption{
\figlabel{fig:srtboth}
$\sqrt{s}$ distributions for baryon baryon collisions
in central  reactions of Au+Au (left)  and 
S+S (right) at  AGS and
SPS energies respectively. 
%}
%\vspace{\cvskip}\caption{\label{fig:srtsps}
%$E^{coll}_{CM}$ distribution for baryon baryon collisions
%in central S+S reaction at SPS, energies. }
%\end{minipage}
%\vspace{\fvskip}\end{figure}
%\vspace{10pt}

\section{Probing the repulsion of the EOS:  flow}

\begin{figure}[htb]
\begin{minipage}[t]{\htwid}
\centerline{\epsfig{figure=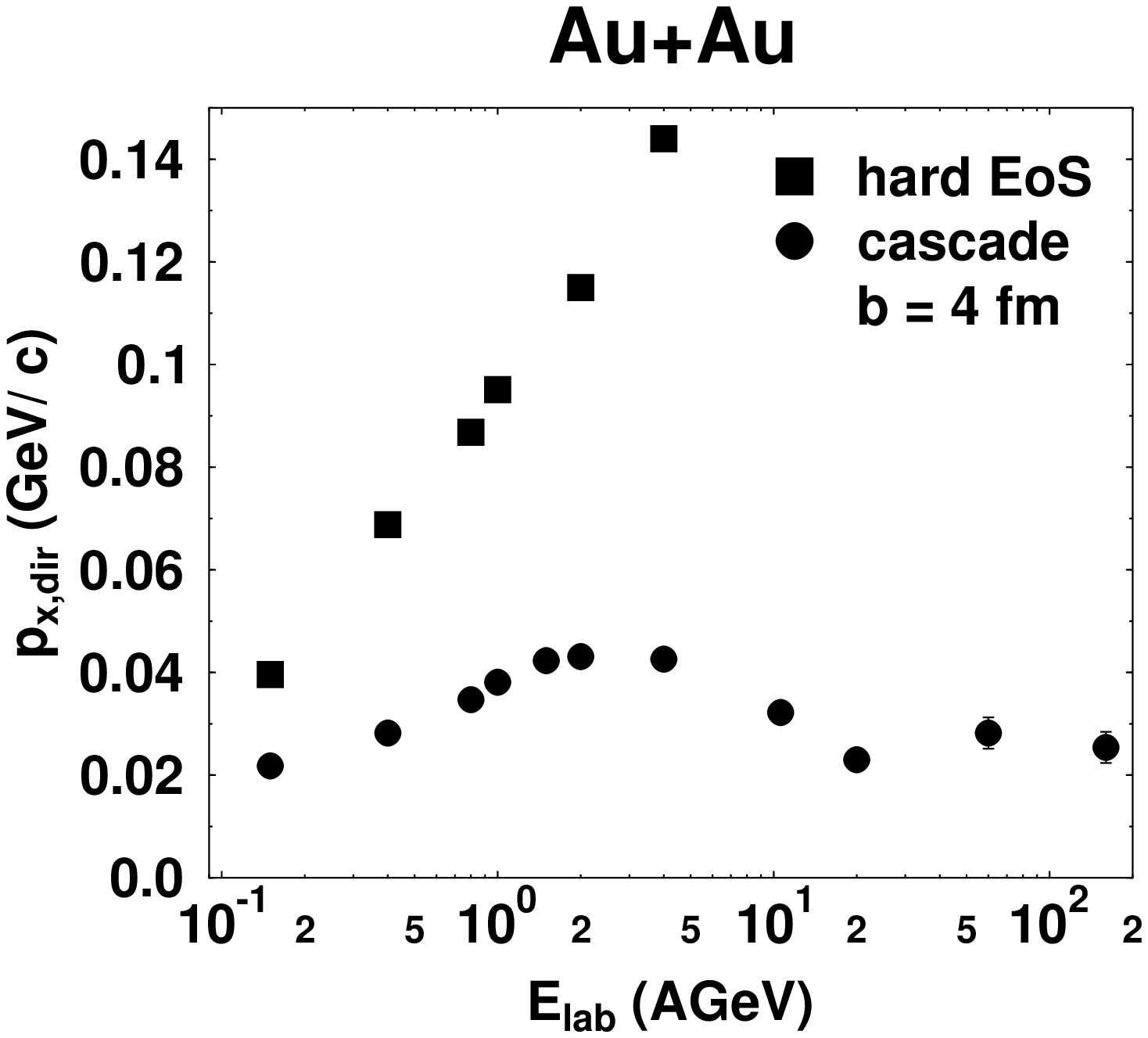,width=\htwid}}
\vspace{\cvskip}\caption{\label{fig:trans1} Excitation function of the total
directed transverse momentum transfer
px-dir for  Au+Au. URQMD calculations including a hard 
EOS (full squares) are compared to the predictions of cascade calculations
(full circles).  }
\end{minipage}
\hfill
\begin{minipage}[t]{\htwid}
\centerline{\epsfig{figure=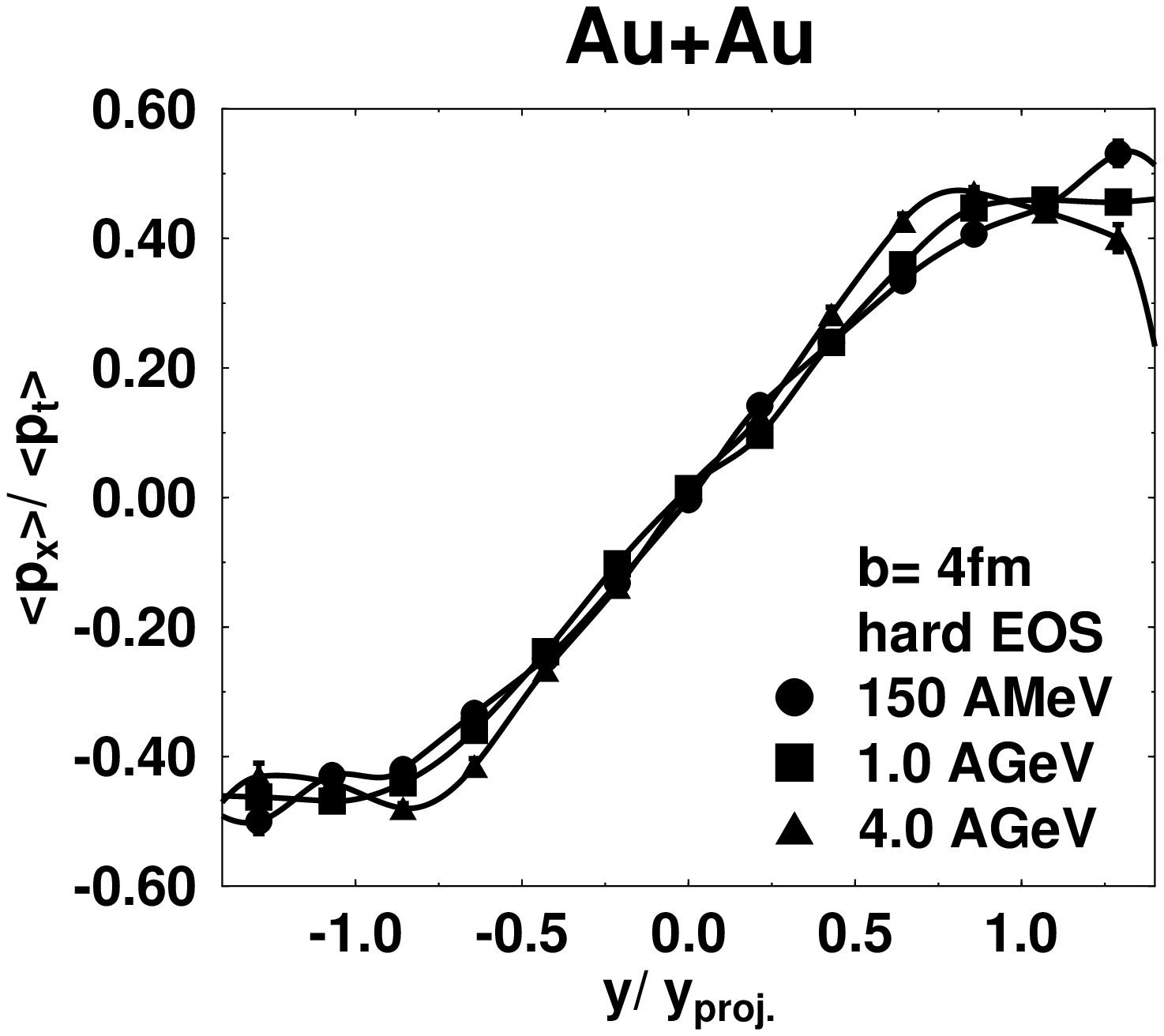,width=\htwid}}
\vspace{\cvskip}\caption{\label{fig:trans2} Mean directed transverse momentum as a
function of the scaled rapidity.
The transverse flow $<p_x>(y/y_{proj.})$  
scales with the mean transverse momentum $<p_t>$,
i.e.\  directivity does not depend on the bombarding energy. }
\end{minipage}
\vspace{\fvskip}\end{figure}

The creation of transverse flow is strongly correlated to the
underlying EOS \cite{Sto86}.
In particular it is believed that
secondary minima as well as the quark-hadron phase transition lead to a
weakening of the collective sideward flow. The occurrence of 
a phase transition should therefore be observable through abnormal
behaviour (e.g.\ jumps) of the strength of collective motion of the
matter \cite{Rischke}.
Note that URQMD  in its present form does not include
any phase transition explicitly.
In Fig.\ref{fig:trans1} the averaged in plane transverse momentum is displayed
for Au+Au from 0.1 to 4~$A$GeV incident kinetic energy.
Calculations employing a hard EOS (full squares) are compared to
cascade simulations (full circles). In the latter case only a slight energy
dependence is observed.
In contrast, the calculation with a hard EOS 
shows strong sensitivity.
%gives a strong dependence.
Here, the integrated
directed transverse momentum  per nucleon is more than twice as high as for
the cascade calculation. This indicates the importance of a non-trivial
equation of state of hadronic matter.

The amount of directed transverse momentum scales in the same way as
the total transverse momentum produced in the course of the reaction. 
Hence, the directivity  depends only on the reaction geometry but not on the
incident energy. This is demonstrateted in  Fig.\ \ref{fig:trans2},
where the mean $p_{\rm x}$ as a function of the rapidity
divided by the average transverse momentum of all particles is plotted.

\section{Temperature dependence of the EOS: photons}

\begin{figure}[htb]
%\vspace*{-2cm}
\begin{minipage}[t]{\htwid}
%\vspace*{-13cm}
%\hspace*{0.2cm}
\centerline{\psfig{figure=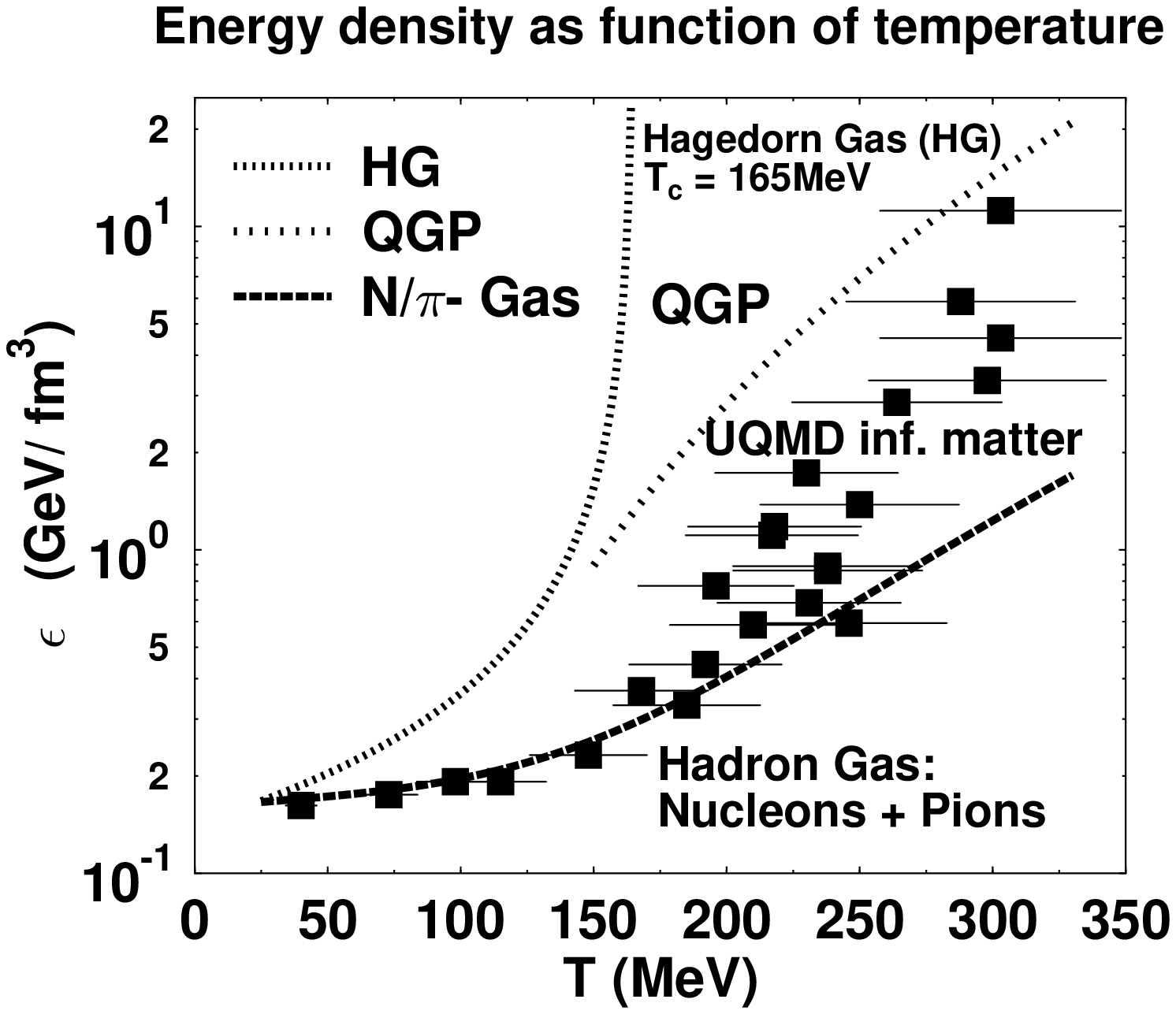,width=\htwid}}
%\end{minipage}
%\hfill
%\vspace*{-5mm}
%\begin{minipage}[c]{\htwid}
%\vspace*{-5mm}
\vspace{0.5\cvskip}\caption{\label{fig:eos}
'EOS' of infinite nuclear matter as a function of
the energy density versus temperature 
at fixed net-baryon
density of $\rho_B=0.16$/fm${}^{3}$ in URQMD (symbols).
%The temperatures resulting from URQMD (symbols)
%are extracted using a least square fit to the baryon momentum spectrum.
The curves refer to analytical forms of the EOS, i.e.\
a Hagedorn-gas (top), a quark-gluon plasma (middle),
and an ideal gas of nucleons and
ultrarelativistic pions (bottom). % are also depicted.
%The solid curve (HG $T_c=300$~MeV) labes an EOS with the
%same functional form as the Hegedorn EOS, but with much higher
%$T_c=300$~MeV.
}
\end{minipage}
\hfill
\begin{minipage}[t]{\htwid}
\centerline{\psfig{figure=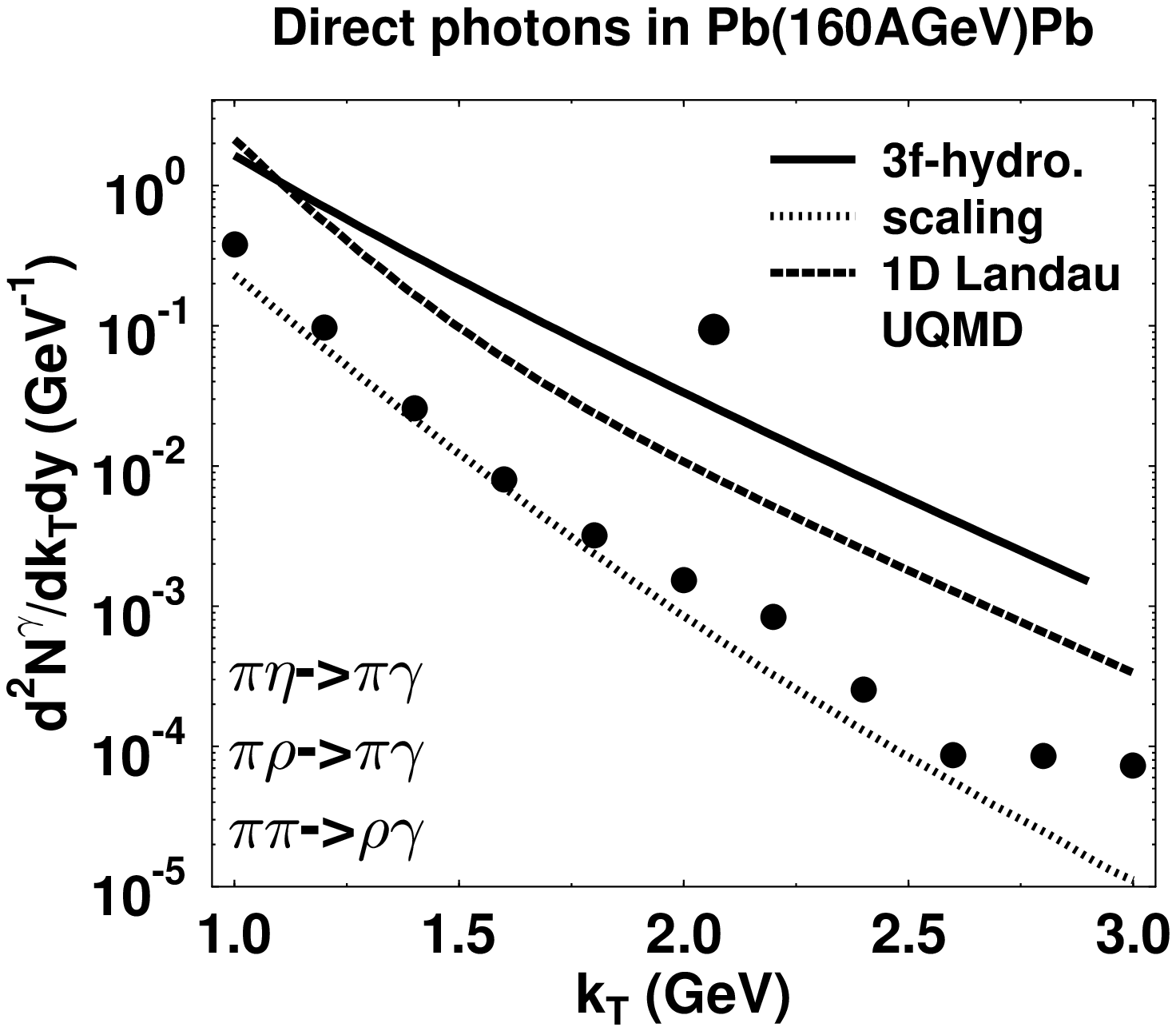,width=\htwid}}
\vspace{0.5\cvskip}\caption{Transverse momentum spectrum of directly produced photons in
Pb+Pb collisions at 160~$A$GeV calculated with URQMD. 
%The contributions of the different processes are shown. 
The resulting spectrum
is compared with hydrodynamical calculations.
In all models the processes 
$\pi\eta \mapsto \pi \gamma$,  $\pi\rho \mapsto \pi \gamma$
and $\pi\pi \mapsto \rho \gamma$ are considered 
as photon sources. 
\label{photon}}
\end{minipage}
\vspace{\fvskip}\end{figure}

%One of the main goals of relativistic heavy ion collisions
%is the determination of the nuclear equation of state.
%In Ref\cite{adrian} photons are prposed to yield
%information about the colling law
%At high energies,
Semiclassical cascade models in terms of scattering hadrons
have proven to be rather
accurate in explaining experimental data.
Therefore it is of fundamental interest to extract
the equation of state from such a microscopic model, i.e.\ to
investigate the equilibrium limits and bulk properties, which are not
an explicit input to the non-equilibrium transport approach with its
complicated collision term (unlike e.g.\  in hydrodynamics\cite{Rischke,adrian}).
In Fig.\ref{fig:eos} the thermodynamic properties of
infinite nuclear matter are studied
within URQMD.

Infinite hadronic matter is simulated in URQMD by constructing %systems
a box of 250~fm$^3$ volume with periodic boundary conditions. % is contructed.
According to the saturation density, 
nucleons are initialized randomly in phase
space, such that a given energy density is reproduced.
%according to a given energy density.
%This corresponds
%ground-state nuclear density.
After the system has equilibrated according to the
simulation with URQMD
the temperature is extracted by fitting the particles'
momentum spectra. Alternatively, the temperature can be extracted from the
relative abundances of different hadrons, e.g.\ the $\Delta/N$ ratio.
%The criterion for thermalisation is choosen as the agreement of the
%temperatures for both methods.

%The result of this procedure is plotted in
%Fig.\ref{fig:eos}. It appears that the
%Temperture dependence much smaller then for the
%Hagedorn gas.
%It yields a 3 times lower energy density
%at $T=150$~MeV as compared to a QGP and does not
%saturates at the Hagedorn limes $T_=165$~MeV.

In Fig.\ref{fig:eos} the result of this procedure is compared
to various analytic forms of the EOS.
While the EOS of a Hagedorn gas and
a QGP yields energy densities $\epsilon \sim 1$GeV/fm${}^3$
at $T=150$~MeV the
temperature dependence is much smaller in URQMD.
It yields about 4-5 times less energy density, being
in fair agreement with a gas composed of nonrelativistic
nucleons and ultrarelativistic pions. It remains
to be seen whether %the disagreement between the Hagedorn model
a reparametrization of
the resonance continuum in the Hagedorn model as suggested in
Ref.\cite{Sto81} would resolve the deviation as compared to URQMD.
On the other hand, beyond $T\sim 200$~MeV
the energy density rises much faster than
$T^4$ approaching even the
QGP value of $\epsilon \sim 10$~GeV/fm${}^3$
around $T=300$~MeV.
This indicates an increase in the number of degrees of freedom.
It may be interpreted as a
consequence of the numerous high mass resonances and string excitations,
which seem to release constituent quark degrees of freedom (but, of
course, no free current quarks as in an ideal QGP).
Investigations
of equilibration times and relative particle and cluster abundances are in
progress. Moreover, the admittedly poor statistics have to be improved,
in order to study the high temperature behavior.

Experimentally, % one can access 
the EOS
can be accessed by measuring electromagnetic radiation
\cite{KLS}. 
In Fig.\ref{photon} the  
direct photon production from meson+meson collisions 
in Pb+Pb collisions at
160~$A$GeV is shown.
Here, only mesons stemming from string decays are included. 
Elastic meson-meson
scattering with $\sigma_{el}=15$mb (independent of
$\sqrt{s}$) was allowed. The result is compared to
calculations within the 3-fluid model \cite{adrian},
scaling and Landau expansion with $T_i=300~$MeV. 
%\bibitem{ADRIAN}A. Dumitru, U. Katscher, J.A. Maruhn, H. St\"ocker,
%W. Greiner, D.H. Rischke: Phys. Rev. C51 (1995) 2166

\section{In medium masses: dileptons}

\begin{figure}[htb]
\begin{minipage}[t]{\htwid}
\centerline{\psfig{figure=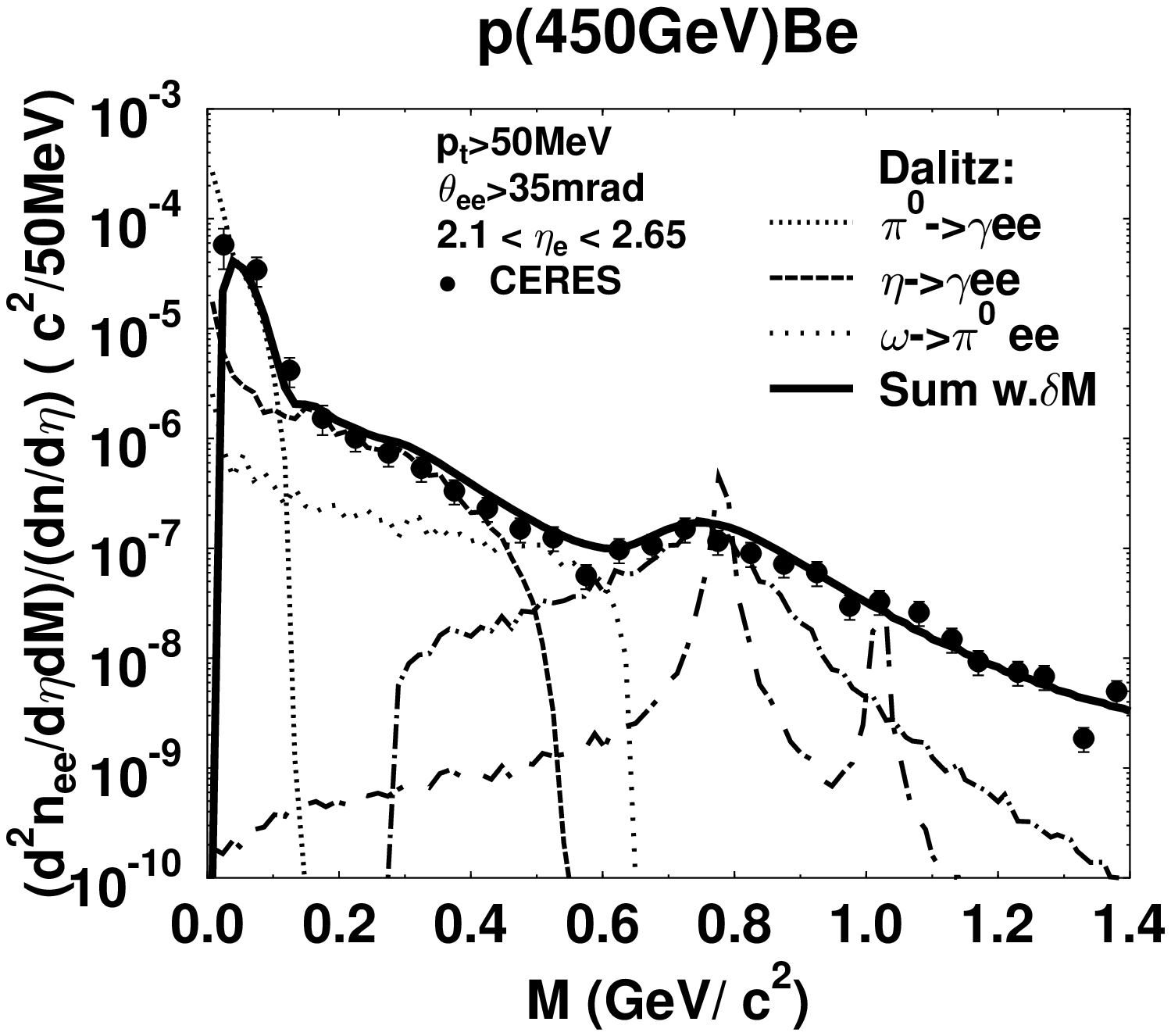,width=\htwid}}
\vspace{\cvskip}\caption{\label{pbe}
Dilepton mass spectrum for $p$+Be at 450~GeV.
The calculation includes Dalitz decays and conversion
of vector mesons (see also legend for S+Au).
The sum of all contributions (solid curve) is folded with the
CERES mass resolution.
%The data are taken from Ref.\cite{specht}.
}
\end{minipage}
\hfill
\begin{minipage}[t]{\htwid}
\centerline{\psfig{figure=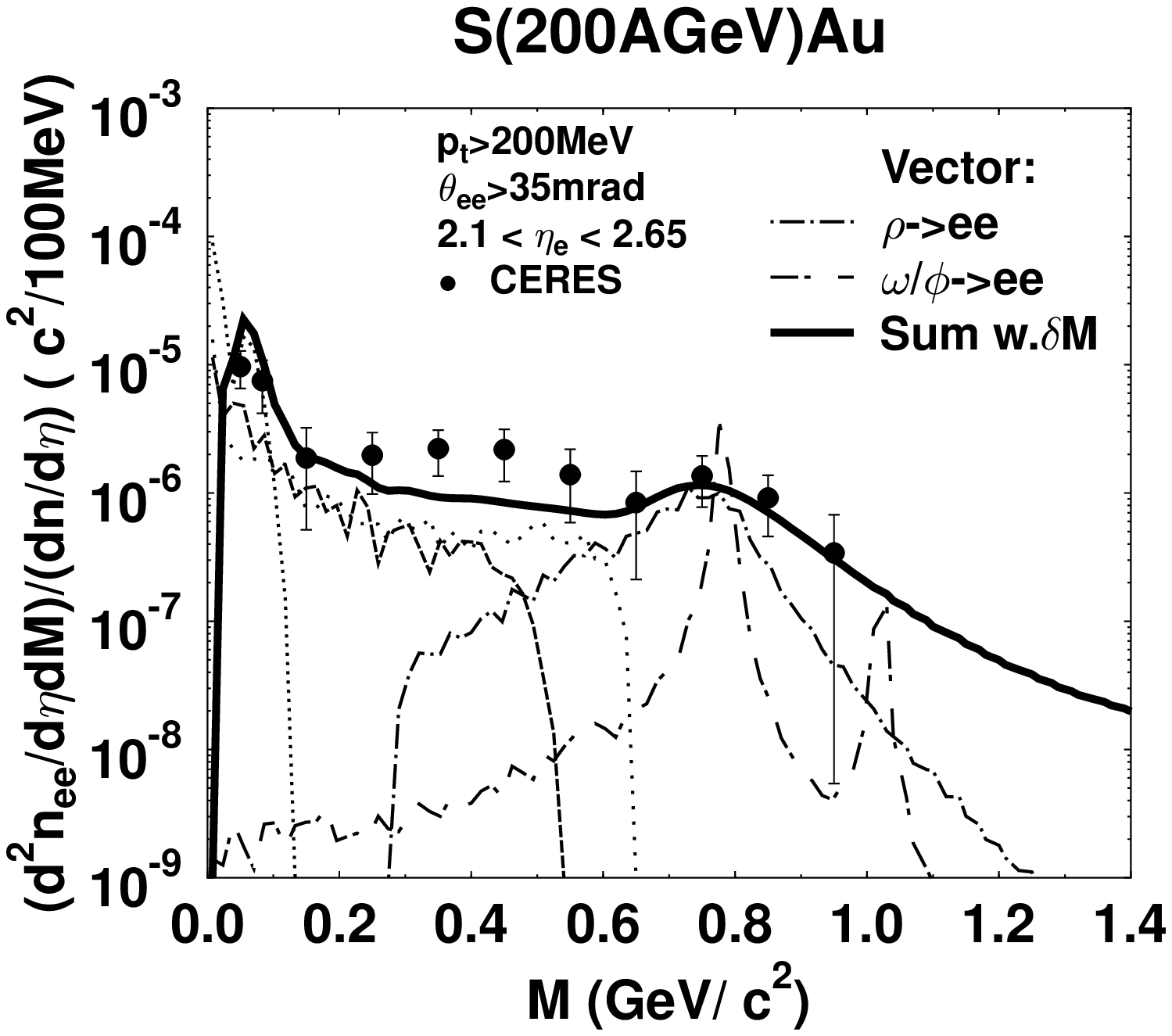,width=\htwid}}
\vspace{\cvskip}\caption{\label{sau}
Dilepton mass spectrum for S+Au at 200~$A$GeV
(see also legend for $p$+Be).
Here no in-medium modifications of the $\rho$ propagator
is considered.
Around $M\sim 400$~MeV two points are missed by $ < 2\sigma$.
%The data are taken from Ref.\cite{specht}.
}
\end{minipage}
\vspace{\fvskip}\end{figure}

In Fig.\ref{pbe} and \ref{sau} calculations of dilepton spectra
with URQMD are shown for $p$+Be
and S+Au. Dilepton sources considered here are
Dalitz decays  ($\pi^0$, $\eta$ and $\omega$)
and vector meson decays ($\rho$, $\omega$ and $\phi$).
Dalitz decays of heavier meson and baryon
resonances are included explicitely via their emission of
$\rho$ mesons  (assuming vector meson dominance).
In order to avoid double counting, the
$\rho$ mesons from $\eta$'s, and $\omega$'s are excluded
from the $\rho$ contribution.
Pion annihilation is included dynamically
into the contribution of decaying $\rho$ mesons
($\pi^+\pi^- \mapsto \rho \mapsto e^+e^-$).
%electromagnetic  bremsstrahlung  is excluded in URQMD.

While the result for $p$+Be agrees well with the
published data from CERES/SPS \cite{specht},  two points around
$M\sim 400$~MeV are missed by about two standard deviations for S+Au.
Speculations about the origin of this deviation include electromagnetic
bremsstrahlung,  annihilations of pions and
a modification of the $\rho$ meson propagator
due to a gradual restoration of the chiral symmetry.

In URQMD the contribution of pion annihilation to the
$\rho$-peak ($\pi^+\pi^- \mapsto \rho$)
is only 40\% for S+Au.
Major additional sources
are decays of  heavy baryons ($\Delta^*/N^* \mapsto N\rho$)
as proposed in Ref.\cite{lw2}
and  meson resonances (see also Fig.\ref{fig:pwid}):
\begin{equation}
\left(\begin{array}{c}  \eta, ~\omega, ~\eta', ~ \phi \\
a_1, ~f_1, ~a_2, ~f_2 \\
\omega(1420), ~\rho(1450) \\ \omega(1600), ~\rho(1700)
\end{array}
\right) \mapsto \left(\begin{array}{c} \rho \gamma \\ \rho \pi \\
\rho \sigma \\ \rho \rho \end{array}
\right)~.
\end{equation}

\begin{figure}[htb]
\begin{minipage}[c]{\htwid}
\centerline{\psfig{figure=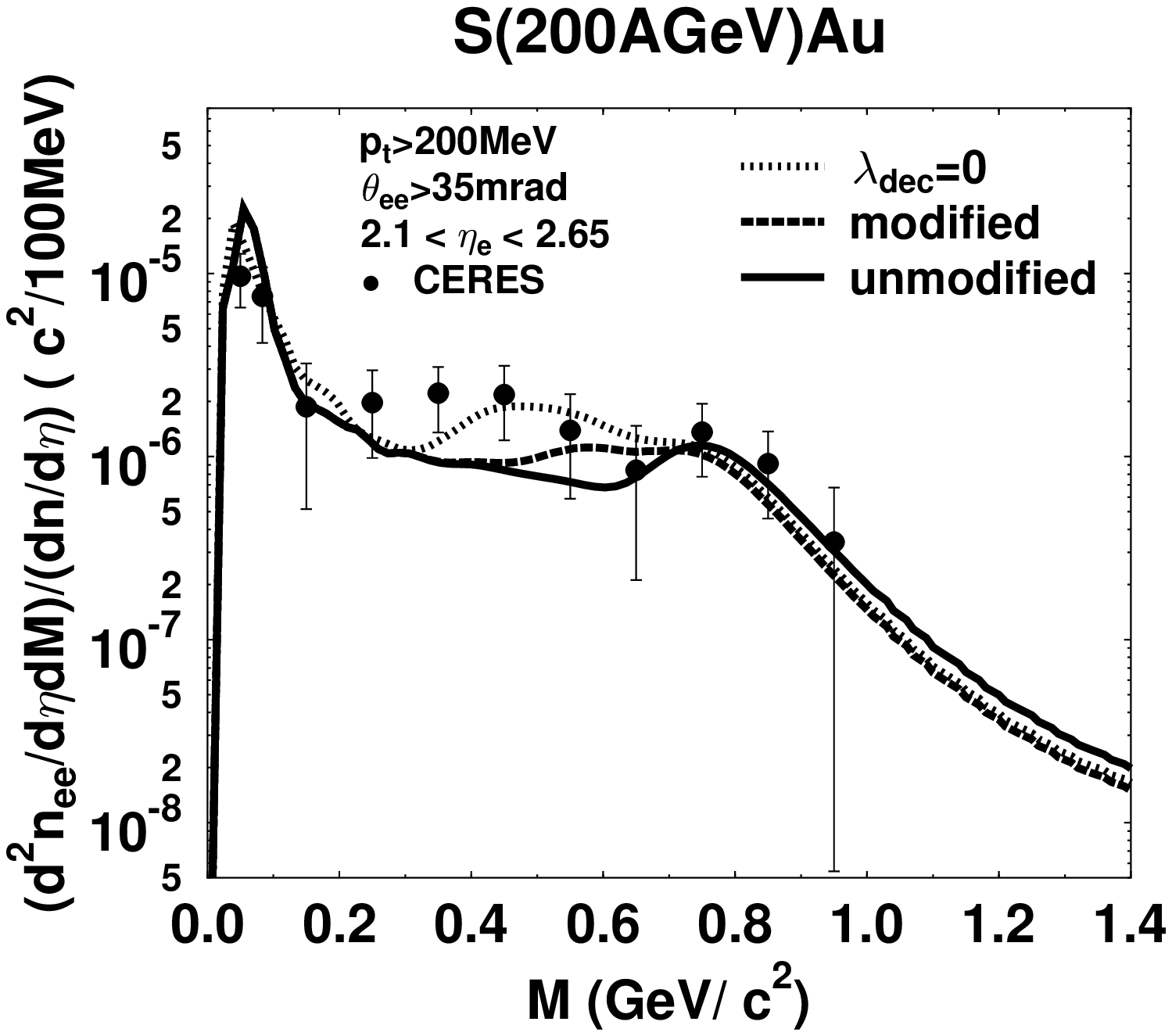,width=\htwid}}
\vspace{\cvskip}\caption
{\label{sau2}
Dilepton mass spectrum for S+Au at 200~$A$GeV.
The curves label simulations with pole shift according to
the creation density (top), the decay density (middle) and without
pole shift (bottom).
%The data are taken from Ref.\cite{specht}.
}
\end{minipage}
\hfill
\begin{minipage}[c]{\htwid}
\centerline{\psfig{figure=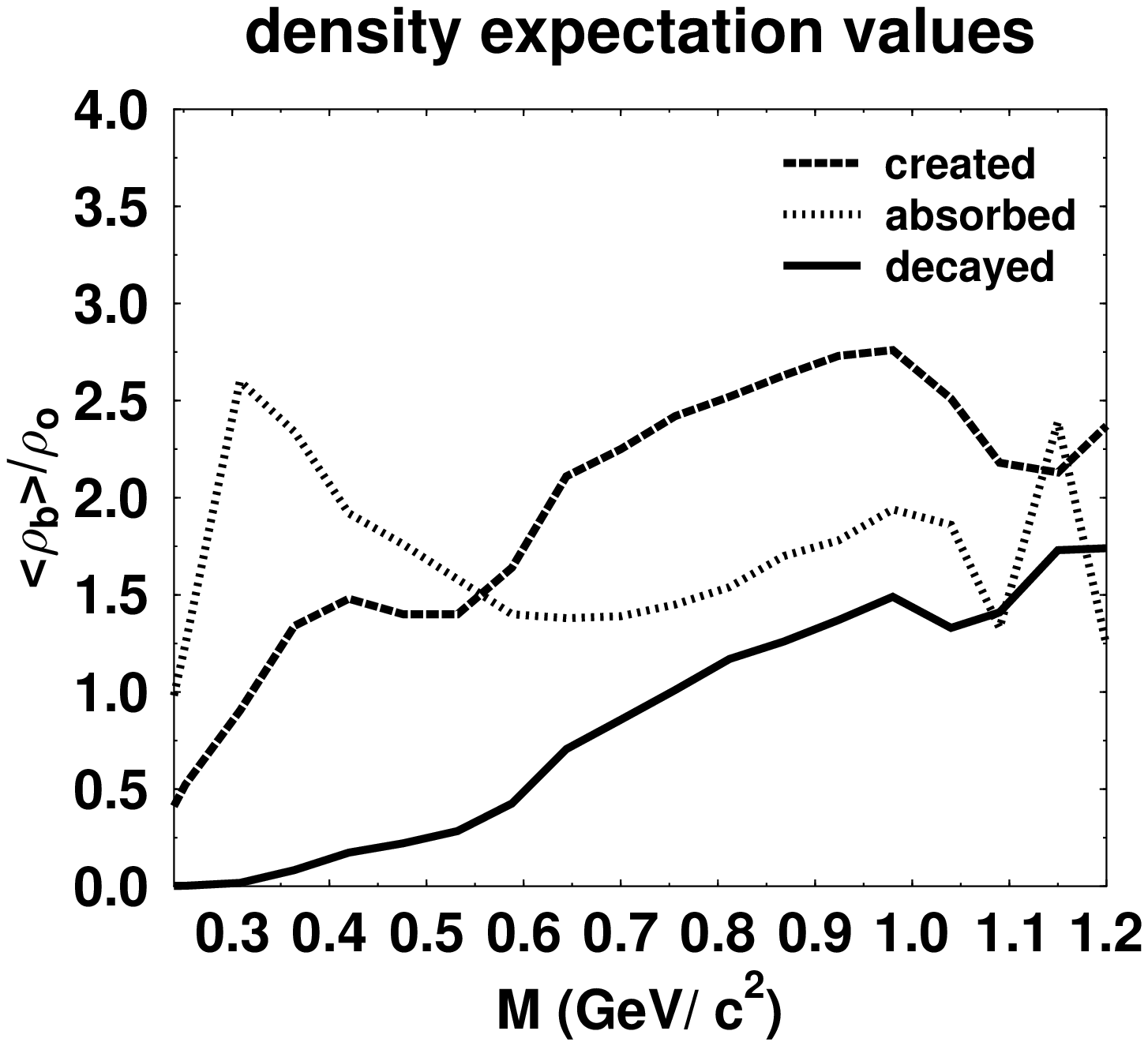,width=\htwid}}
\vspace{\cvskip}\caption
{\label{rhoabx}
Dependency of $\rho$-interactions on the density expectation values. The curves
show the values for the creation, absorption and decay of $\rho$ mesons. 
Note that $\langle \varrho_b\rangle < 0.5 \varrho_{0}$ 
around $M\simeq 0.4$~GeV.
}
\end{minipage}
\vspace{\fvskip}\end{figure}
In Ref.\cite{GEB91,THa92a} a
linear dependence of the  $\rho^0 / \omega $ pole position
as a function  of the nuclear density  $\rho $  has been  suggested:
 $m_{\rho^0}(\varrho/\varrho_0) 
	= m_{\rho^0} (0) ( 1- \lambda \varrho/\varrho_0)$.
Here  $\rho_0 $ denotes the ground state density of nuclear matter, and
 $\lambda=0.18 $, in agreement with various other calculations.
Since the restriction to low densities may not be suitable
for heavy ion collisions, the following extrapolation towards higher
densities  is  taken:
\be
    m_{\rho^0}(\varrho/\varrho_0)
     = \frac{m_{\rho^0} (0)}{ 1 + \lambda ~ \varrho/\varrho_0}
    \label{mrho}
    ~.
\ee
In Fig.\ref{sau2} an application of
Eq.(\ref{mrho}) is made to calculate
a dielectron mass spectrum for a  density dependent
vector meson pole.
This result yields only a small enhancement around $M \sim 500$~MeV
as compared to the calculation without pole shift (bottom curve).
On the other hand, the data can nicely be reproduced,
if the strong (unphysical) assumption is made, that the
pole at the decay point ($\rho \mapsto e^+e^-$)
is shifted according to the creation density (upper curve).
This would be a neglection of the finite decay length.
The discrepancy of a calculation without 
decay length ($\lambda_{\rm dec}=0$~fm) as compared to the result 
including the decay length is driven by two reasons:
i) The increase of the $\rho$ lifetime ($\sim 7$~fm/$c$) below 
its resonance mass 
in the region $M\sim 0.3-0.5$~GeV (where
a dilepton excess in S+Au is reported\cite{specht})
lowers the decay density down
to $\langle \varrho \rangle \sim 0.2 \varrho_o$ for S+Au (or $0.3\varrho_o$ for
Pb+Au).
ii) An enhancement of the decay length leads to an increase of 
reabsorption. Hence, the radiation path for 
$\rho \mapsto e^+e^-$ is substantially truncated. This fact is further
investigated in Fig.\ref{rhoabx}, where the effect of the mass dependent
$\rho$-width is depicted by the course of the density dependence for $\rho$
mesons with invariant masses $M$. The decreasing width (i.e. increasing
lifetime) of low-mass-resonances leads to higher mean decay times where the
baryon density is already dilute. Thus in-medium-corrections can only yield 
small enhancements when treated this way. 
However, the interpretation of the experimental data 
gives the following impression: The data for light systems such as 
$p+$Be and $p+$Au (see also Ref.\cite{ko,cassing})
as well as the data for the heavy Pb+Au system both for inclusive   and central
reactions are reproducible without mass shifts. In contrast, the 
central data for S+Au exceed the URQMD calculation
around $M\sim 0.4$~GeV by about two standard deviations.

\begin{figure}[htb]
\begin{minipage}[c]{1.2\htwid}
\centerline{\psfig{figure=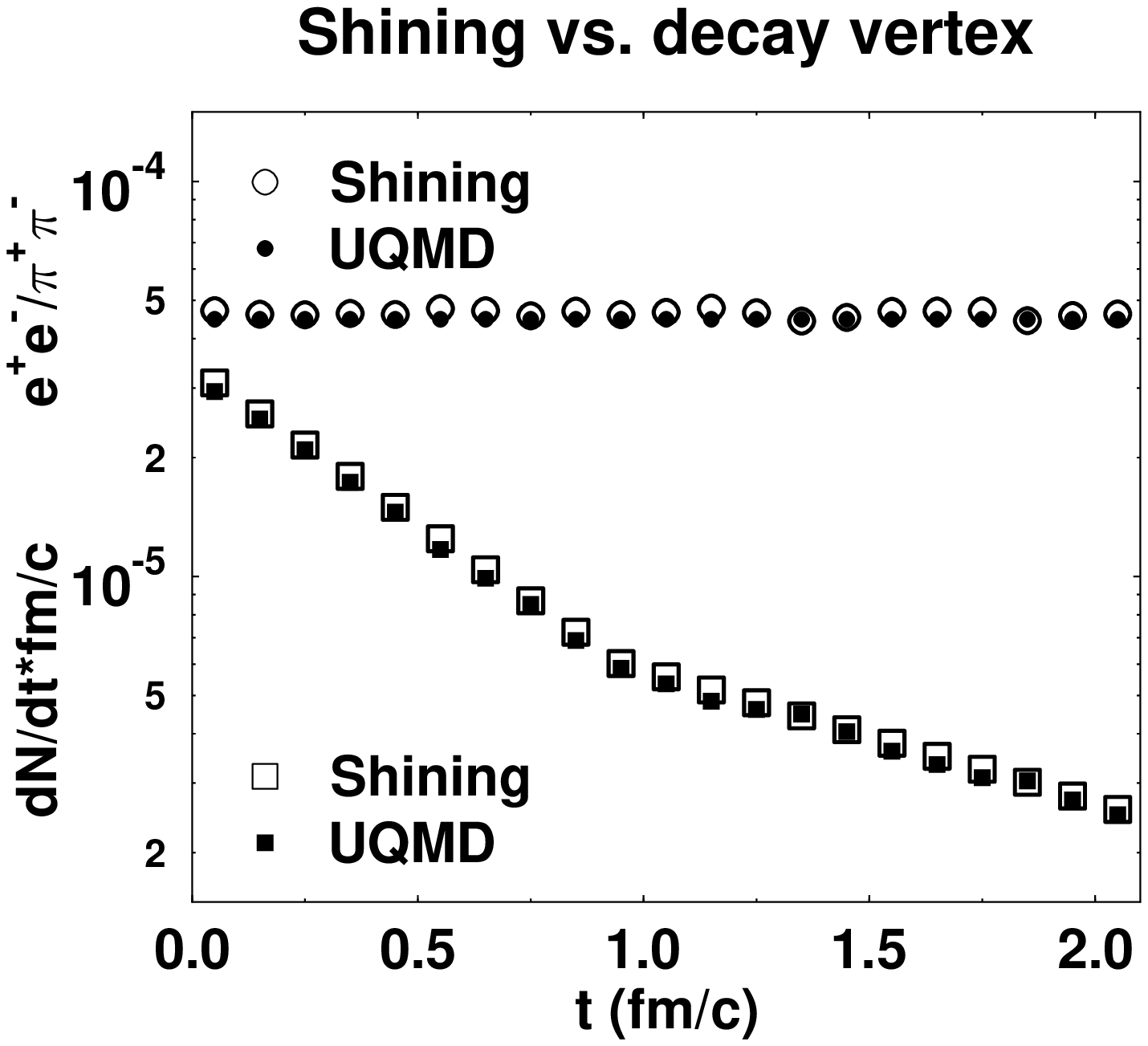,width=\htwid}}
%\vspace{\cvskip}
\end{minipage}
\hfill
\begin{minipage}[c]{0.8\htwid}
\caption{\label{dec}
Comparision of the treatment of shining $\rho$-sources to the URQMD
approach,
where the dileptons are created at the $\rho$ decay vertex. 
%A time dependent baryon density is modelled by 
For $t<1$~fm/$c$ a constant
$\rho$ absorption is considered.
Both methods show the same time dependence.
}
\end{minipage}
\vspace{\fvskip}\end{figure}

\begin{figure}[htb]
\begin{minipage}[c]{\htwid}
\centerline{\psfig{figure=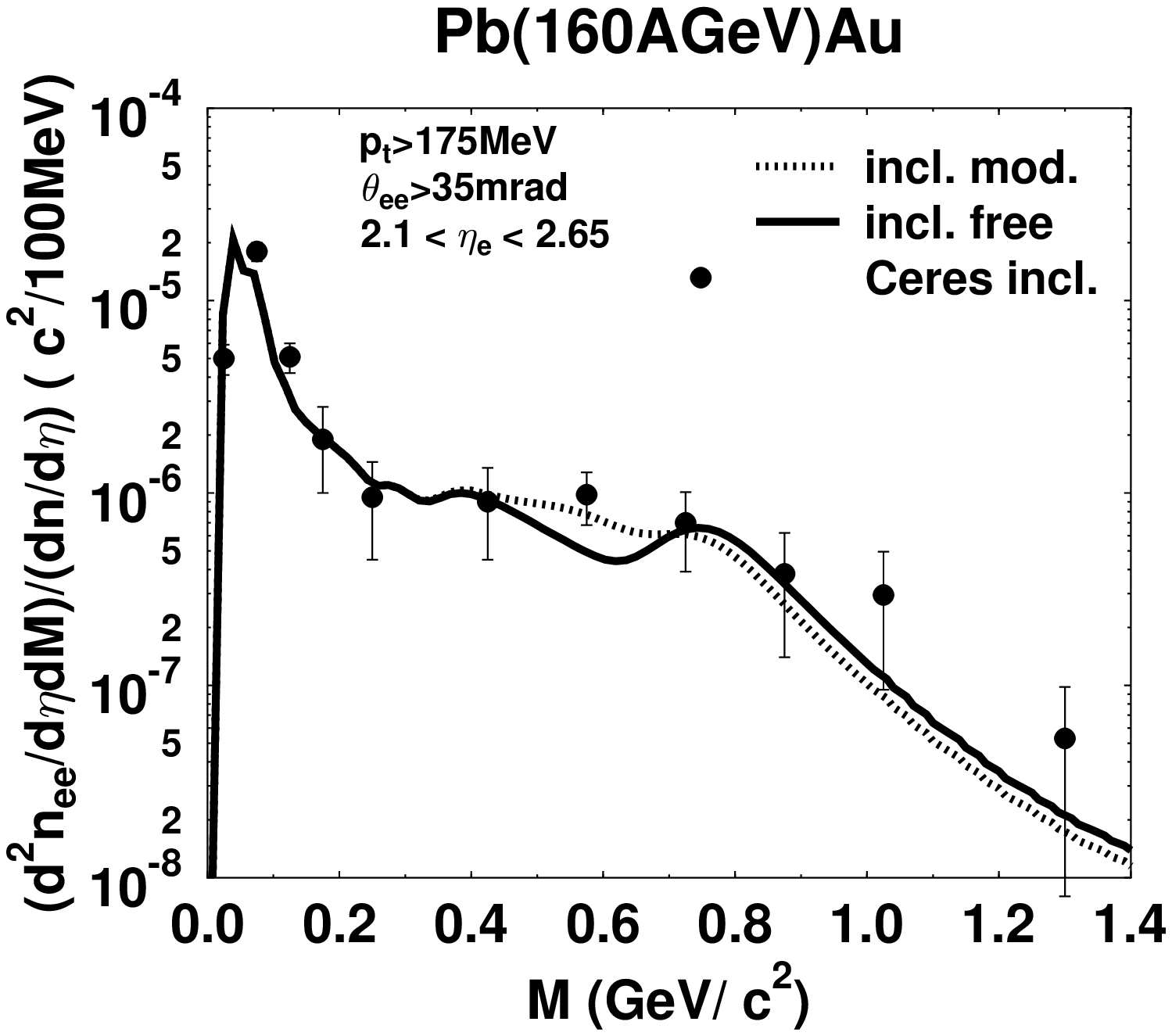,width=\htwid}}
\vspace{\cvskip}
\caption{\label{pbau}
Dilepton mass spectra for Pb+Au at 160~$A$GeV.
The curves label calculations 
for inclusive reactions  with (mod.) and
without a pole shift (free).
%The  bottom curve shows the result for peripheral events without pole shift.
The symbols refer to preliminary data for peripheral events
from CERES.
}
\end{minipage}
\hfill
\begin{minipage}[c]{\htwid}
\centerline{\psfig{figure=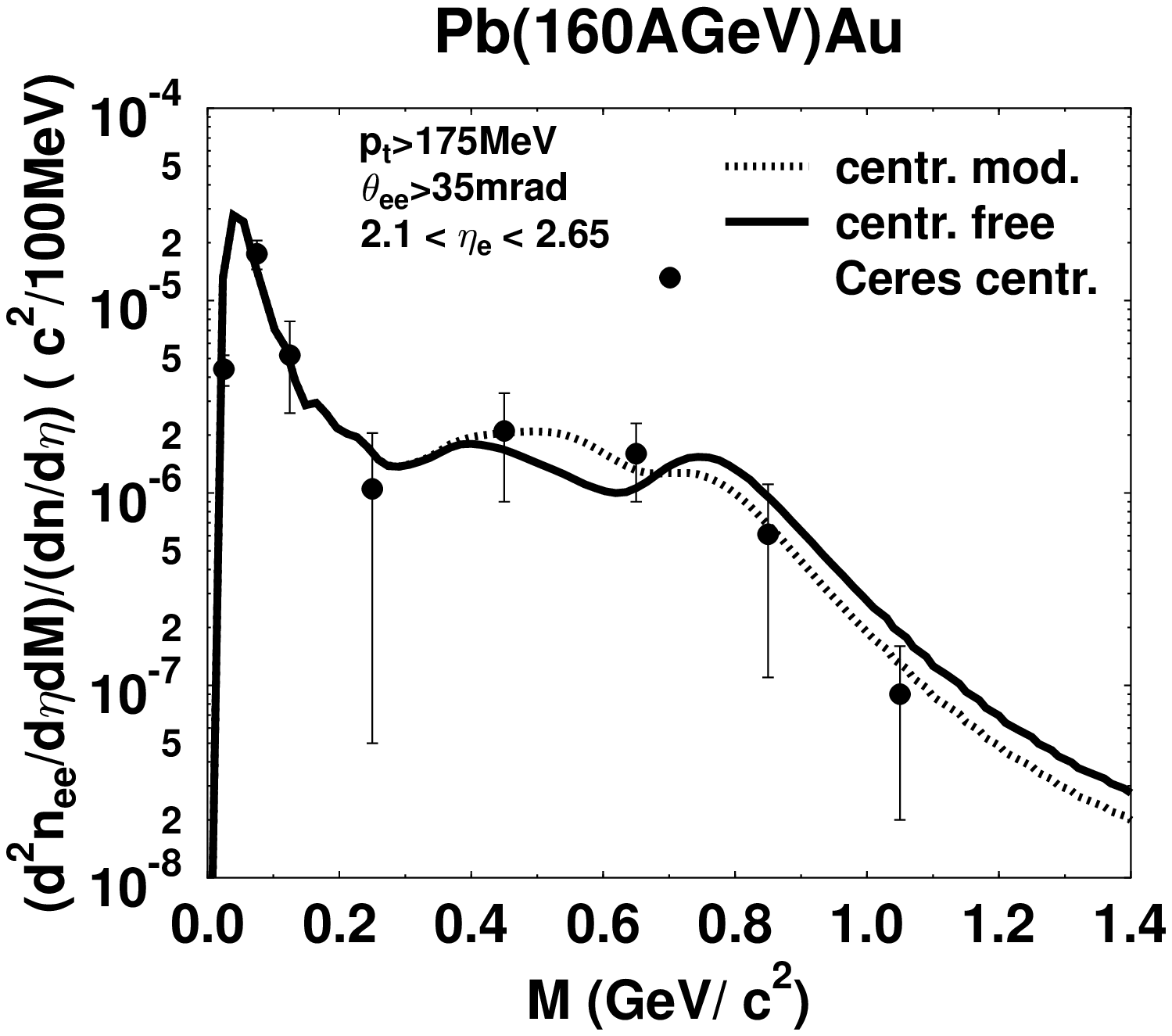,width=\htwid}}
\vspace{\cvskip}
\caption{\label{pbauc}
Dilepton mass spectra for Pb+Au at 160~$A$GeV.
The curves label calculations 
for central 
reactions  with (mod.) and
without a pole shift (free).
%The  bottom curve shows the result for peripheral events without pole shift.
The symbols refer to preliminary data for high multiplicity events
from CERES.
}
\end{minipage}
\vspace{\fvskip}\end{figure}

In URQMD the $\rho$ meson pole position is shifted
according to the density at which the $\rho$ meson decays,
i.e.\ eventually converts
into $e^+e^-$. % (middle curve). 
Note that this procedure
is equivalent to  a ''shining''
description,  where the $\rho$ constantly emits $e^+e^-$ pairs
according to the rate $dN^{ee}/dt = \Gamma (\rho \mapsto e^+e^-)$.
In Fig.\ref{dec} the shining description is compared to the treatment in URQMD
for a $\rho$ source at rest including a time dependent 
reabsorbation probability. %of 50$\%$ up to 1~fm/$c$. 
Both methods yield - whithin the statistical limits - 
the same average emission times. Furthermore a simple time dependent
density profile is used to calculate the mean $e^+e^-$ emission density.
Again, both methods yield the same values. 

The results for Pb+Au are shown in Figs.\ref{pbau} and \ref{pbauc}.
Both calculations for inclusive reactions
are in fair agreement with
the preliminary observation from CERES\cite{drees}.
The result for central events is given in Fig.\ref{pbauc}. 
%yields an  enhancement due to
%a nonlinear in-medium $\rho$-decay contribution
%The $\rho$ bump is located either at
%$\sim 500$~MeV with pole shift or at $\sim 700$~MeV without
%modification.
Note that the calculations without modifications
are well compatible for $p$+Be and Pb+Au for both centralities.
Only in S+Au reactions two datapoints are missed by the default calculation
by about two standard
deviations. 
%However, the changes induced by the pole shifts as compared
%to the calculation without pole shift 
%are for both systems and centralities
%lower than the statistical errors.

\section{Summary}

Studies of the equation of state and consequences
of gradual restoration of the chiral symmetry
are presented using a novel microscopic phase space model, URQMD,
including 75 hadron species and strings.
The directed tranverse momentum 
shows strong 
sensitivities to the underlying EOS:
It is small in the cascade calculation,
whereas it scales linearly with the
average transverse momentum for a hard equation of state.
Hence, measureing the excitation function of
the transverse directed flow allows for a systematic study
of the EOS.
The calculation of dilepton yields without modifications of the $\rho$ mass
pole is well compatible with the CERES-data for $p$+Be and Pb+Au. Only in S+Au
reactions two datapoints at $0.3-0.5$~GeV are missed
by the default URQMD-calculation by about two standard deviations. 
%The increase of the $\rho$ lifetime  at
%masses where
%a dilepton excess in S+Au is reported\cite{specht}
%lowers its average $e^+e^-$ emission density substantially. 
%Therefore it seems that the density dependence 
%of the $\rho$ pole
%does not suffice to explain the dilepton excess in S+Au.
%Hence, other effects  -- such as the temperature dependence
%of \qq{} or additional sources -- 
%%and an improvement of the statistics 
%might be required.

\setlength{\listparindent}{2cm}

\end{document}